\documentclass[
reprint,
superscriptaddress,
amsmath,
amssymb,
aps,
longbibliography,
pra,
showpacs,
floatfix
]{revtex4-1}

\usepackage{graphicx}
\usepackage[table]{xcolor}

\usepackage{tabularx}
\usepackage{multirow}
\usepackage{dcolumn} 
\newcolumntype{d}[1]{D{.}{.}{#1}}
\newcolumntype{L}[1]{>{\raggedright\arraybackslash}p{#1}}
\newcolumntype{C}[1]{>{\centering\arraybackslash}p{#1}}
\newcolumntype{R}[1]{>{\raggedleft\arraybackslash}p{#1}}
\usepackage{booktabs}

\usepackage{comment}
\usepackage{enumitem}

\newcommand{\ang}{\ensuremath{\mathrm{\AA}}}

\usepackage[hidelinks]{hyperref}
\definecolor{color1}{rgb}{0,0.25,0.70}
\hypersetup{colorlinks=true, 
    linkcolor={color1},
    citecolor={color1}, 
    urlcolor={color1}
}

\begin{document}

\title{Highly confined phonon polaritons in monolayers of oxide perovskites}

\author{Dominik~M.\ Juraschek}
\email{djuraschek@seas.harvard.edu}
\affiliation{Harvard John A. Paulson School of Engineering and Applied Sciences, Harvard University, Cambridge, MA 02138, USA}

\author{Prineha Narang}
\email{prineha@seas.harvard.edu}
\affiliation{Harvard John A. Paulson School of Engineering and Applied Sciences, Harvard University, Cambridge, MA 02138, USA}

\date{\today}

\begin{abstract}
Two-dimensional (2D) materials are able to strongly confine light hybridized with collective excitations of atoms, enabling electric-field enhancements and novel spectroscopic applications. Recently, freestanding monolayers of oxide perovskites have been synthesized, which possess highly infrared-active phonon modes and a complex interplay of competing interactions. In this study, we evaluate central figures of merit for phonon polaritons in the tetragonal phases of the 2D perovskites SrTiO$_3$, KTaO$_3$, and LiNbO$_3$, using density functional theory calculations. Specifically, we compute the 2D phonon-polariton dispersions, the propagation-quality, confinement, and deceleration factors, and we show that they are comparable to those found in the prototypical 2D dielectric hexagonal boron nitride. Our results suggest that monolayers of oxide perovskites are promising candidates for polaritonic platforms in the terahertz spectral range that enable possibilites to control complex phases of matter through strongly enhanced electromagnetic fields.
\end{abstract}

\maketitle


\section{Introduction}

\begin{figure}[b]
\centering
\includegraphics[scale=0.102]{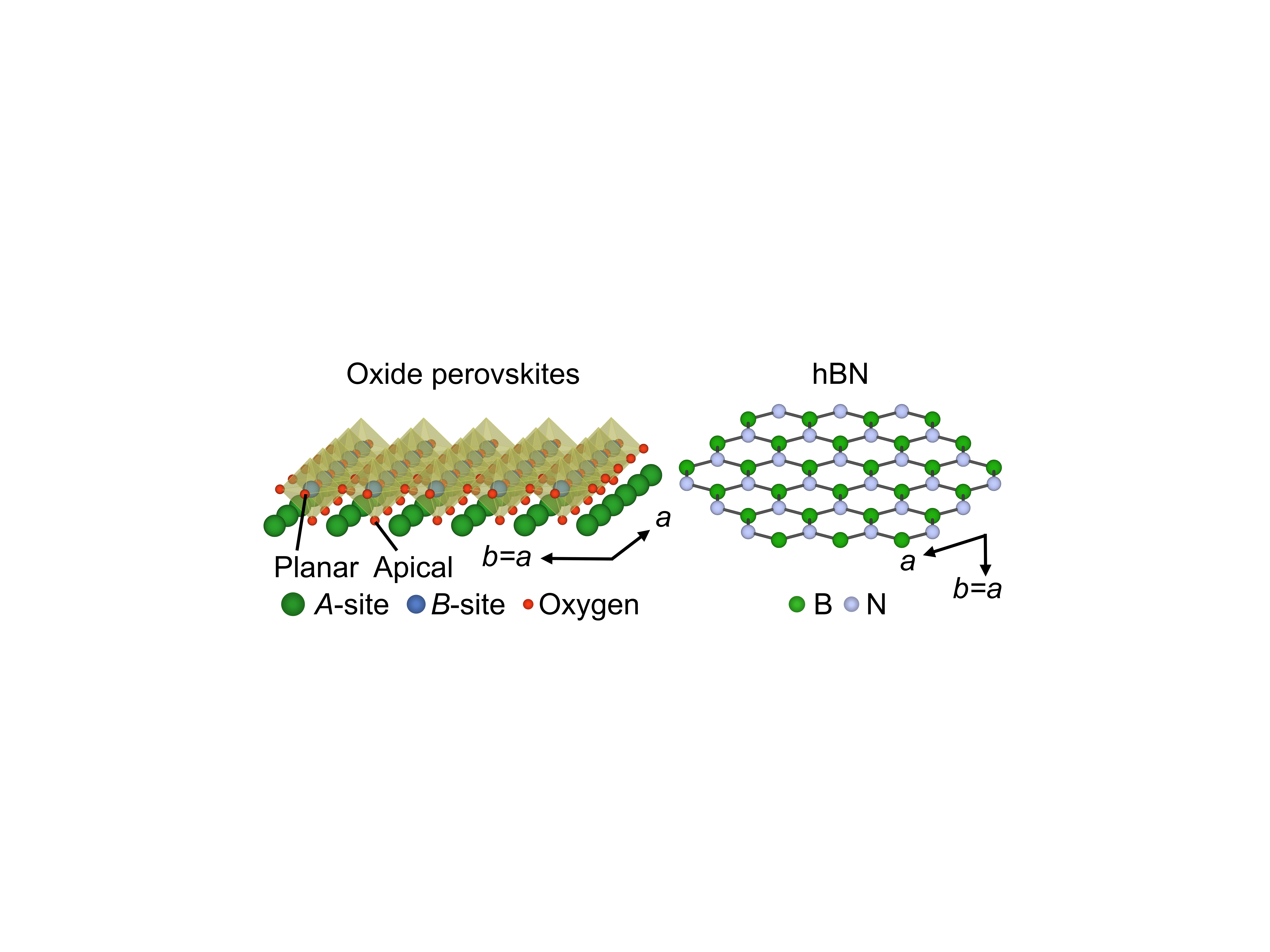}
\caption{
Structures of tetragonal $AB$O$_3$ perovskite monolayers and hexagonal boron nitride. The oxygen octahedra around the transition-metal $B$-site cations are incomplete and miss the top apical oxygen ions.
}
\label{fig:2dstructure}
\end{figure}

Phonon polaritons are hybrid excitations of light and collective atomic vibrations that couple the characteristics of photons to the length scales and nonlinearities intrinsic to the crystal lattice. As a result, the wavelength of light in the polariton quasiparticle is compressed to scales much smaller than in free space, enabling novel spectroscopic methods beyond the diffraction limit \cite{Basov2016,Low2017,Basov2021}, enhanced nonlinear light-matter interactions \cite{Reshef2019,Rivera2020}, and enhancements of dipolar emission \cite{Rivera2016,Rivera2017}. In low-dimensional and 2D materials, polaritons are easily accessible by nanoscale near-field imaging and the reduced dimensionality leads to additional confinement of electromagnetic energy \cite{Basov2016,Low2017}. A prototypical example is hexagonal boron nitride (hBN) (Fig.~\ref{fig:2dstructure}), in which phonons within the hexagonal layers couple to light in the mid-infrared spectral region \cite{Dai2014,Caldwell2014,Shi2015,Yoxall2015,Woessner2015,Dai2015,Caldwell2019}. In monolayers of hBN, these phonon polaritons strongly confine light and reductions of the wavelength by factors of $>300$ with respect to free space have recently been achieved \cite{Dai2019,Li2021}.

At the same time, the class of 2D materials has consistently been growing, involving more and more systems beyond van der Waals materials \cite{Osada2019,Gibertini2019,Guan2020}. Particularly intriguing materials are oxide perovskites, which are known to display rich phenomena in their bulk phases, including high-temperature superconductivity \cite{Bednorz1988}, metal-insulator transitions \cite{Imada/Fujimori/Tokura:1998}, and multiferroicity \cite{Hill:2000}. Recently, freestanding membranes of oxide perovskites have been created \cite{Lu2016,Hong2017,Chen2019,Lu2019,Dong2019,Harbola2020,Bourlier2020,Sun2020_2,Wang2020,Dong2020,Gu2020,Xu2020,Cai2020,Hong2020,Han2020,Li2020_9,Lee2020}, whose thicknesses reach down to the monolayer limit \cite{Ji2019}, and which are highly tunable with homogeneous and inhomogeneous strain \cite{Lu2018,Xu2020,Hong2020,Han2020,Zhang2020,Xu2020_9,Cai2020,Lee2020}. Because oxide perovskites conventionally possess large effective ionic charges \cite{Rabe2007}, they are promising candidates for light-matter interactions in the 2D limit.


Here, we investigate phonon-polaritonic properties of oxide-perovskite monolayers and benchmark them against those in well-known hBN. Specifically, we use density functional theory to compute the 2D phonon-polariton dispersions and polaritonic figures of merit, including propagation-quality, confinement, and deceleration factors, of the highly dispersive high-frequency phonons in three $AB$O$_3$-type perovskites: SrTiO$_3$, KTaO$_3$, and LiNbO$_3$. We chose these candidates, because they are well studied in literature and at the same time continuously new phenomena and applications are being discovered and predicted, such as multiferroic quantum criticality \cite{Narayan2019}, engineering of superconductivity \cite{Ahadi2019} and ferroelectricity \cite{Nova2019,Li2019,Mankowsky:2017,VonHoegen2018,Abalmasov2020}, nonlinear optical and electro-optical applications \cite{Wang2018_2,Zhang2019}, novel Hall effects \cite{Li2020}, and even the generation of Moir\'{e} lattices \cite{Burian2020}. We find that while the propagation quality of 2D phonon polaritons in the oxide perovskites is slightly lower than that found in hBN due to the complexity of the perovskite structure, their confinement and deceleration factors yield comparable numbers. With these properties, oxide perovskites broaden the spectral range of polaritonic platforms within the terahertz regime.


\section{Theoretical formalism}

We begin by reviewing the theoretical formalism for phonon-polaritonic figures of merit in 2D materials. The derivations of the equations used for these figures of merit are presented in detail in the works of Sohier et al. and Rivera et al.~\cite{Sohier2017,Rivera2018,Rivera2019}, and in the Appendix. The three key figures of merit that we investigate here are the propagation-quality, confinement, and deceleration factors. The propagation-quality factor, $Q$, is a measure for the distance (in numbers of wavelengths) that the phonon polariton travels before it decays and is defined by the ratio of the imaginary to the real part of the lattice optical conductivity, $\sigma$. The confinement factor, $C$, is a measure for the compression of the wavelength of light trapped in a phonon polariton with respect to its wavelength in free space and is defined by the ratio of the phonon-polariton wavevector, $q$, to that of free light, $q_0$. The deceleration factor, $D$, is a measure for the slowdown of light when trapped in a phonon polariton and we define it as the ratio of the group velocity, $v_g=\partial_q \omega$ ($\omega$ being the wavevector-dependent frequency of the phonon polariton), to the speed of light, $c$. The three factors read
\begin{equation}\label{eq:figuresofmerit}
Q=\frac{\mathrm{Im}\sigma}{\mathrm{Re}\sigma}, ~~ C=\frac{q}{q_0}, ~~ D=\frac{\partial_q \omega}{c}.
\end{equation}

In order to evaluate the above equations, we need to compute the lattice optical conductivity and the dispersion of the 2D phonon polaritons. In conventional 3D bulk systems, phonon polaritons form bands around the residual-ray region between the transverse-optical (TO) and longitudinal-optical (LO) phonon frequencies. In 2D however, the reduced dimensionality of the Coulomb interaction does not result in a splitting between LO and TO phonons at long wavelengths, instead their branches converge at the center of the Brillouin zone ($\Gamma$ point) \cite{Sohier2017,Rivera2019}. The generalization of the nonanalytical-term correction describing the relation between LO and TO frequencies, $\omega_{\alpha,\mathrm{LO}}$ and $\omega_{\alpha,\mathrm{TO}}$, across dimensions reads
\begin{equation}\label{eq:LOTO}
\omega_{\alpha,\mathrm{LO}}^2 - \omega_{\alpha,\mathrm{TO}}^2 = V(q) \frac{q^2}{\Omega}|\hat{\mathbf{q}}\cdot\mathbf{Z}_\alpha|^2.
\end{equation}
Here, $\hat{\mathbf{q}}$ is the unit wavevector (propagation direction), $\mathbf{Z}_\alpha$ is the mode effective charge vector of phonon mode $\alpha$, and $\Omega$ is either the unit-cell volume (3D) or area (2D). The mode effective charge vector is given by $\mathbf{Z}_\alpha=\sum_{n} Z^\ast_n \hat{\mathbf{e}}_{\alpha n}/\sqrt{M_n}$, where $Z^\ast_n$ is the Born effective charge tensor of atom $n$, $\hat{\mathbf{e}}_{\alpha n}$ is the phonon eigenvector, $M_n$ is the atomic mass, and the index $n$ runs over all atoms in the unit cell \cite{Gonze1997}. $V(q)$ is a dimensionality-dependent screened Coulomb interaction and in 2D given by $1/(2q\varepsilon_0\varepsilon_\mathrm{2D})$, where $\varepsilon_\mathrm{2D}$ is the dielectric function of the 2D material.

While a full 2D treatment of LO/TO phonons is implemented in current first-principles packages \cite{Sohier2017}, a simple analytical formula can be obtained taking into account a linear dependence of the 2D dielectric function and the surroundings of the 2D material (e.g. vacuum for a freestanding monolayer), $\varepsilon_\mathrm{2D}=\varepsilon_\mathrm{env}+qr_\mathrm{eff}=1+qr_\mathrm{eff}$, where $r_\mathrm{eff}$ is the effective screening length. The dispersion of the 2D LO phonon is given by \cite{Rivera2019}
\begin{equation}\label{eq:dispersionomega}
\omega_{\alpha,\mathrm{LO}}^2 = \omega_{\alpha,\mathrm{TO}}^2 + \frac{Nq\mathcal{S}_\alpha}{1+Nqr_\mathrm{eff}},
\end{equation}
where $N$ is the number of layers of the material ($N=1$ for a monolayer), and $\mathcal{S}_\alpha=|\hat{\mathbf{q}}\cdot{}\mathbf{Z}_\alpha|^2/(2\Omega\epsilon_0)$. Here, the second term on the right hand side vanishes for $q\rightarrow 0$ and the LO and TO branches merge at the $\Gamma$ point. It has previously been shown that the 2D LO phonon branch fulfills exactly the conditions necessary for a highly confined evanescent mode in a dielectric monolayer, similar to a surface phonon polariton in 3D \cite{Rivera2019}. The 2D LO phonon can therefore be regarded as the 2D phonon polariton, similar to the equivalence between 2D plasmons and plasmon polaritons. We will accordingly use the notions ``2D LO phonon'' and ``2D phonon polariton'' interchangeably in the remainder of the manuscript.

The lattice optical conductivity can be computed with knowledge of the phonon frequencies and Born effective charges. In the long-wavelength limit ($q\rightarrow 0$), $\sigma$ can be expressed as
\begin{equation}\label{eq:opticalconductivity}
\sigma(\omega) = -2i\varepsilon_0\omega\sum\limits_{\alpha} \frac{\mathcal{S}_\alpha}{\omega_\alpha^2 - \omega^2 - 2i\gamma_\alpha\omega},
\end{equation}
where $\omega_\alpha$ is the phonon frequency at the $\Gamma$ point and $\gamma_\alpha$ is the phonon linewidth. Only infrared-active phonon modes possess nonzero mode effective charges and therefore contribute to the ionic (lattice) part of the optical conductivity.


\begin{table*}[t]
\centering
\def\arraystretch{1.25}
\caption{
Calculated in-plane lattice constants, $a$, $\Gamma$-point phonon frequencies, $\omega_\alpha/(2\pi)$, phonon linewidths at room temperature, $\gamma_\alpha$, and in-plane mode and Born effective charges, $|\mathbf{Z}_\alpha|$ and $Z^*_i$. $e$ is the elementary charge and amu the atomic mass unit. We distinguish between apical (ap) oxygen ions and displacements parallel and perpendicular to the $B$-O bond for planar (p) oxygen ions. We further show the fitted effective screening lengths, $r_\mathrm{eff}$, and effective layer number, $N_\mathrm{eff}$.
}
\begin{tabularx}{\linewidth}{XXXXXXXXXXXX}
\hline\hline
Material & $a$ [\ang{}] & $\omega_\alpha/(2\pi)$ [THz] & $\gamma_\alpha$ [THz] & $|\mathbf{Z}_\alpha|$ [$e/\sqrt{\mathrm{amu}}$] & $Z^*_A$ [$e$] & $Z^*_B$ [$e$] & $Z^*_{\mathrm{O,ap}}$ [$e$] & $Z^*_{\mathrm{O,p}||}$ [$e$] & $Z^*_{\mathrm{O,p}\perp}$ [$e$] & $r_\mathrm{eff}$ [\ang{}] & $N_\mathrm{eff}$\\
\hline
SrTiO$_3$ & 3.83 &  15.7 & 0.18 & 1.0 & 2.5 & 7.1 & -1.8 & -5.8 & -2.0 & 6.0 & 2.0 \\
KTaO$_3$ & 3.86 & 17.3 & 0.28 & 1.3 & 1.2 & 7.9 & -1.3 & -6.2 & -1.6 & 8.7  & 1.5 \\
LiNbO$_3$ & 3.85 & 17.5 & 0.093 & 1.0 & 1.0 & 5.9 & -0.5 & -4.6 & -1.8 & 5.3 & 1.4 \\
hBN & 2.51 & 41.0 & 0.035 & 1.1 & 2.7 & -2.7 & & & & 7.3 &  \\
\hline\hline
\end{tabularx}
\label{tab:properties}
\end{table*}


\begin{figure}[b]
\centering
\includegraphics[scale=0.0925]{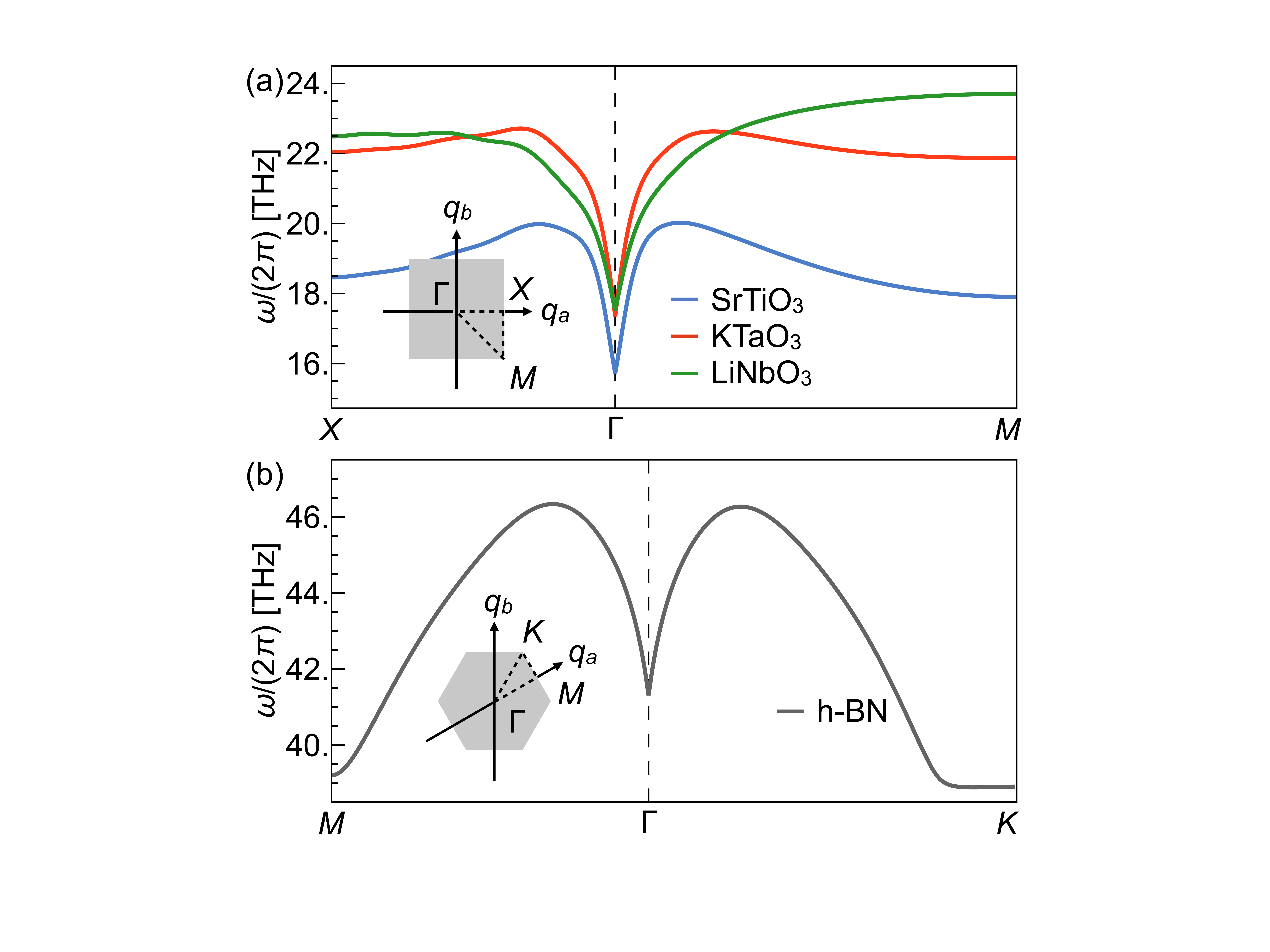}
\caption{
Wavevector dependence of the 2D LO phonon frequencies, $\omega/(2\pi)$, along high-symmetry lines in reciprocal space. Shown are the high-frequency $E$ modes in (a) monolayers of oxide perovskites and in (b) hexagonal boron nitride. Insets show the Brillouin zone paths. In 2D, there is no LO-TO splitting at the $\Gamma$ point.
}
\label{fig:dispersions}
\end{figure}


\section{Results}

We now turn to the evaluation of the first-principles calculations and the polaritonic figures of merit. Computational details are provided in the Appendix. We use tetragonal symmetry for the unit cells of the oxide-perovskite monolayers, which have been found and predicted for various compounds \cite{Ji2019,Xiao2019,Xu2020_9}. We show the in-plane lattice constants of our fully relaxed tetragonal structures (space group $P4mm$, no. 140) and of hBN (space group $P6_3/mmc$, no. 194) in Table~\ref{tab:properties}. The values for the oxide perovskites are slightly lower (by $\sim0.5\%{}$) than those obtained in a previous density functional theory study using hybrid functionals \cite{Xiao2019}. We did not find any imaginary frequencies in the calculation of the phonons, which indicates that the tetragonal structures are indeed stable.

For the oxide perovskites, we focus on the highest-frequency degenerate in-plane phonon modes (irreducible representation $E$), because their branches do not cross any other phonon bands over the entire Brillouin zone path. hBN only possesses one pair of degenerate in-plane $E$-symmetric phonon modes. We display the values for the calculated frequencies, linewidths, and mode effective charges of these phonon modes in Table~\ref{tab:properties} together with the Born effective charges of the ions. In Fig.~\ref{fig:dispersions}, we show the calculated wavevector dependence of the highest-frequency 2D LO phonons/2D phonon polaritons for the oxide perovskites and hBN along high-symmetry lines in reciprocal space. Phonons in both materials show highly dispersive behavior near the $\Gamma$ point, whereas the phonon bands in the oxide perovskites flatten out for large wavevectors in contrast to hBN. LiNbO$_3$ is the only of the investigated compounds, whose phonon frequency nearly monotonically increases with increasing wavevectors.

\begin{figure}[t]
\centering
\includegraphics[scale=0.095]{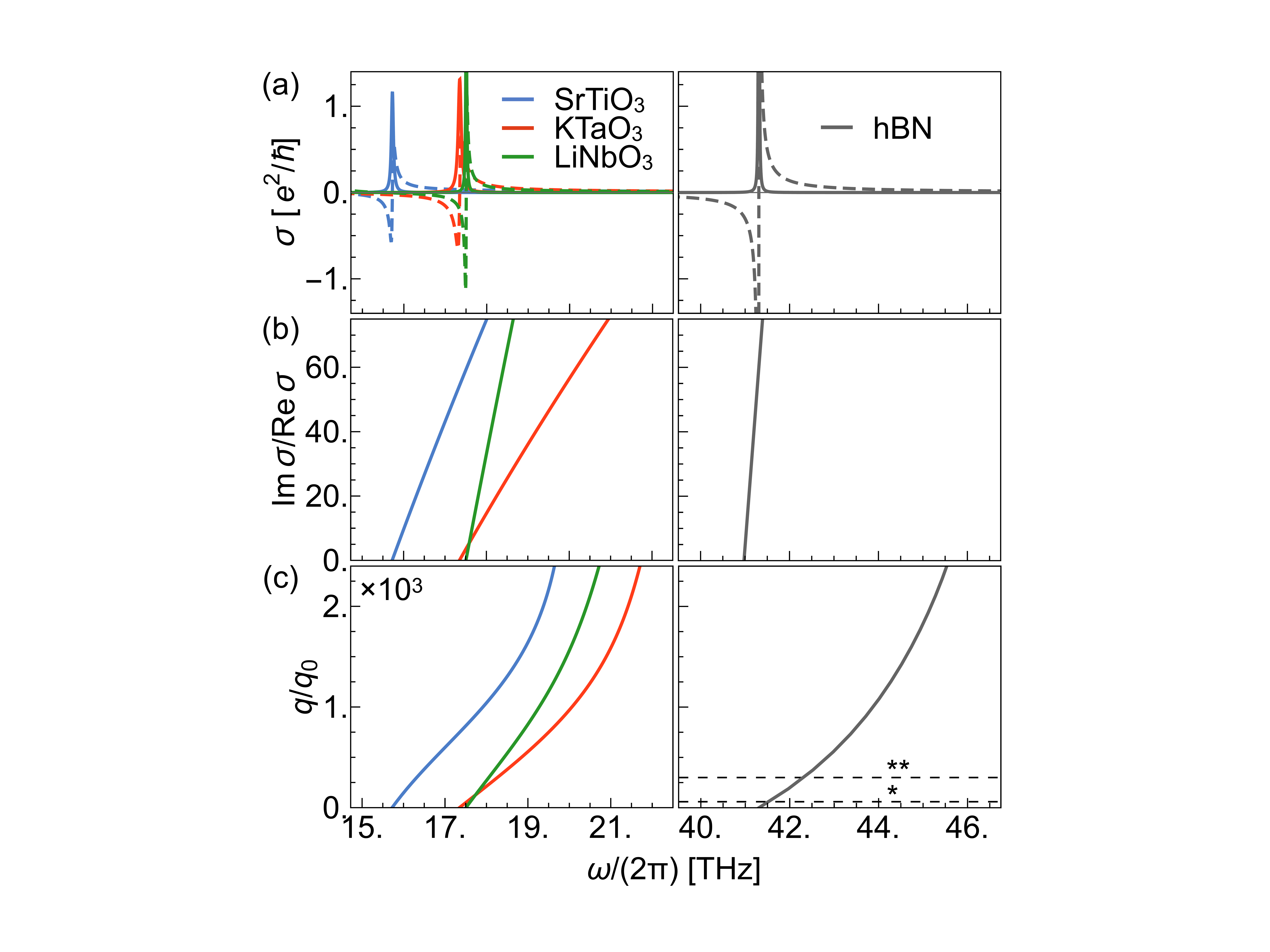}
\caption{
Polaritonic figures of merit for monolayer oxide perovskites (left) and hexagonal boron nitride (right). (a) Real (solid lines) and imaginary (dashed lines) parts of the lattice optical conductivity, Re$\sigma$ and Im$\sigma$, for the high-frequency doubly degenerate infrared-active $E$ modes. (b) Propagation-quality factor, $Q=\mathrm{Im}\sigma/\mathrm{Re}\sigma$, measuring the number of wavelengths that the 2D phonon polariton travels before decaying. (c) Confinement factor, $C=q/q_0=\lambda_0/\lambda$, measuring the wavelength of the 2D phonon polariton, $\lambda$, compared to that of free light, $\lambda_0$. Dashed lines denote confinement factors in hBN achieved in topical measurements, *\cite{Dai2019} and **\cite{Li2021}.
}
\label{fig:figuresofmerit}
\end{figure}

In Fig.~\ref{fig:figuresofmerit}(a), we show the real and imaginary parts of the lattice optical conductivities, Re$\sigma$ and Im$\sigma$, computed using Eq.~(\ref{eq:opticalconductivity}). The parameters determining the shape of the lattice optical conductivities are the mode effective charges, the area of the unit cell, and the phonon linewidths. While the mode effective charges of the oxides perovskites and hBN are nearly equal (1-1.3~$e/\sqrt{\mathrm{amu}}$ compared to 1.1~$e/\sqrt{\mathrm{amu}}$ for hBN), the oxide perovskites have roughly 2-3 times larger unit-cell areas and at the same time 3-8 times larger $\Gamma$-point phonon linewidths (Table~\ref{tab:properties}). These latter two factors make the peak magnitudes of the lattice optical conductivities smaller for the oxide perovskites than for hBN. The magnitude of the phonon linewidths is further the main determining parameter for the propagation-quality factors of the 2D phonon polaritons, as described by Eq.~(\ref{eq:opticalconductivity}), which we show in Fig.~\ref{fig:figuresofmerit}(b). The propagation quality factors, $Q=\mathrm{Im}\sigma/\mathrm{Re}\sigma$, for the oxide perovskites show a slower increase with increasing frequency (and therefore wavevector) than the one of hBN due to their larger phonon linewidths. LiNbO$_3$ displays the largest lattice optical conductivity, as well as the highest propagation quality of the three oxide perovskites, with a slope similarly steep as hBN, while the slopes for SrTiO$_3$ and KTaO$_3$ are notably flatter.


\begin{figure}[t]
\centering
\includegraphics[scale=0.095]{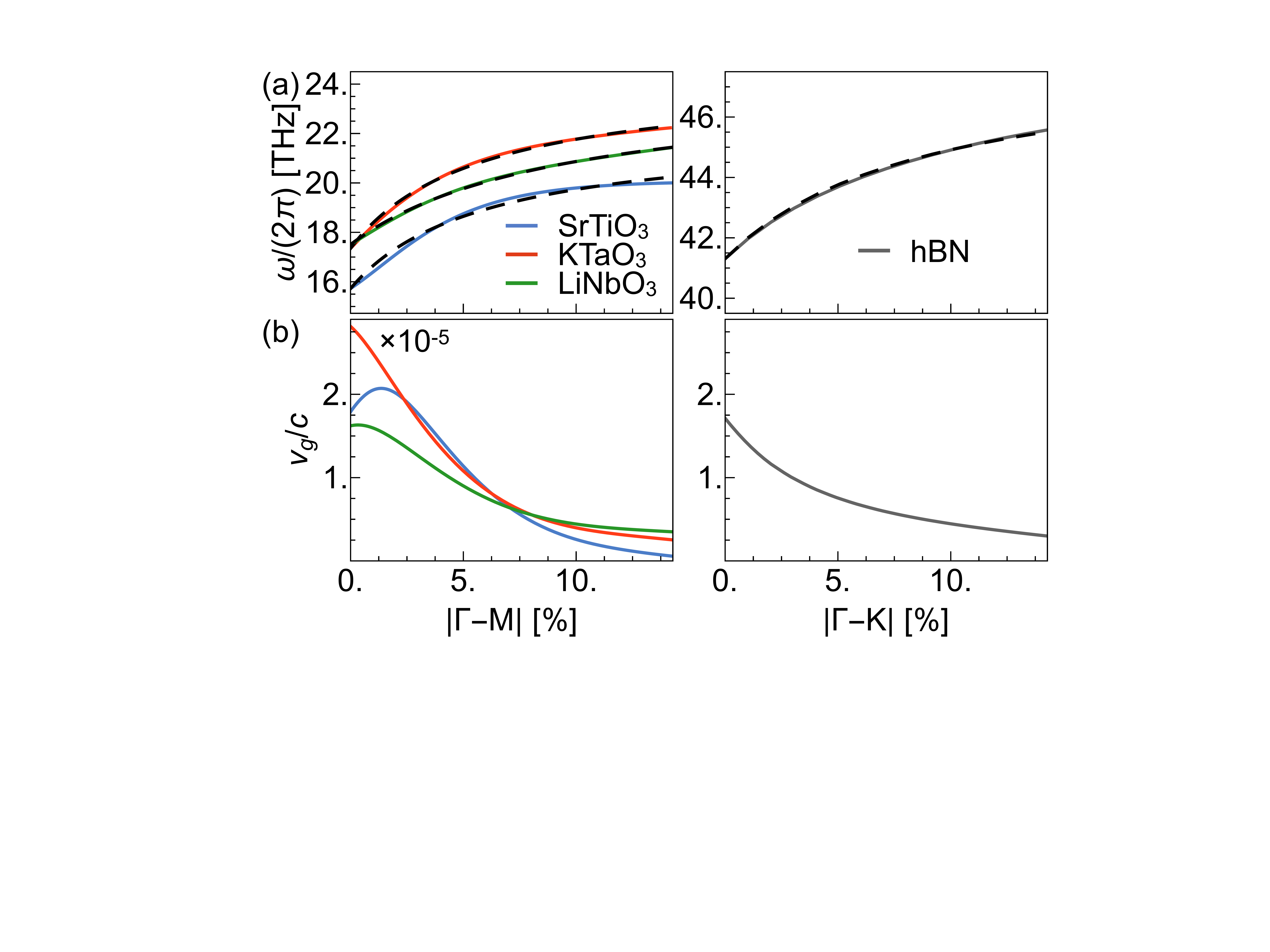}
\caption{
2D phonon-polariton dispersions and slowdown of light in the monolayer oxide perovskites (left) and hexagonal boron nitride (right). (a) Comparison of the ab initio dispersions (solid lines) with the model from Eq.~(\ref{eq:dispersionomega}) (dashed lines). Shown are the data for the high-frequency $E$ modes for a portion of the $\Gamma$-$M$ and $\Gamma$-$K$ directions, respectively. (b) Deceleration factor, $D=v_g/c=\partial_q \omega/c$, measuring the slowdown of light trapped in the 2D phonon polaritons with respect to free space.
}
\label{fig:dispersionvelocity}
\end{figure}

We have evaluated the lattice optical conductivities and propagation-quality factors in the long-wavelength limit above. In the following, we use the explicit wavevector dependence of the 2D phonon polaritons to evaluate the confinement factors according to Eq.~(\ref{eq:figuresofmerit}), which we show in Fig.~\ref{fig:figuresofmerit}(c). Both confinement factors for the oxide perovskites and hBN show a similarly steep increase with increasing frequency and reach the order of several thousands before the peak and subsequent flattening of the 2D phonon-polariton dispersion is reached. Dashed lines mark the confinement factors achieved in hBN in topical experiments, which reached $C\sim 60$ in a recent near-field microscopy study \cite{Dai2019} and $C>300$ in a recent electron energy-loss spectroscopy study \cite{Li2021}. Those numbers are still well within the linear regime of the 2D phonon-polariton dispersions and therefore confinement factors in Fig.~\ref{fig:figuresofmerit}(c), which suggests that much larger values can possibly be achieved in both the oxide perovskites and in hBN the future.

We next look into the behavior of the 2D phonon polaritons close to the $\Gamma$ point. In Fig.~\ref{fig:dispersionvelocity}(a), we show a portion of the phonon dispersion from Fig.~\ref{fig:dispersions} of around 15\%{} of the $\Gamma$-$M$ and $\Gamma$-$K$ directions in reciprocal space for the oxide perovskites and hBN, respectively. We fit Eq.~(\ref{eq:dispersionomega}) to the ab initio wavevector dependence over these portions of the high-symmetry lines in order to extract the effective screening length, $r_\mathrm{eff}$. For hBN, the value that we obtain, $r_\mathrm{eff}=7.3~\ang{}$, is reasonably close to the $7.6~\ang{}$ found in Ref.~\cite{Sohier2017_2} using ultrasoft pseudopotentials. For the oxide perovskites however, we are not able to properly fit the ab initio wavevector dependence by optimizing only $r_\mathrm{eff}$ in Eq.~(\ref{eq:dispersionomega}). Instead, we find that fitting $r_\mathrm{eff}$ and an effective layer number, $N\equiv N_\mathrm{eff}$, produced good results, which suggests that the structure of the oxide perovskites is complex enough that even monolayers possess multilayer character to a certain extent. We display the effective screening lengths and layer numbers in Table~\ref{tab:properties}.

In the following, we investigate the slowdown of light in the 2D phonon polaritons, as described by the deceleration factor in Eq.~(\ref{eq:figuresofmerit}). We show the deceleration factors for the different compounds in Fig.~\ref{fig:dispersionvelocity}(b) for the same portion of the high-symmetry lines as above. The group velocities of the 2D phonon polaritons in the oxide perovskites are of the same order of magnitude as that of hBN, slowing down the propagation of light by a factor of 1-3$\times$10$^{-5}$ with respect to the speed of light in vacuum, $c$. At 7.5\%{} of the respective high-symmetry lines, the group velocities of all investigated compounds reach a similar value of around 0.6$\times$10$^{-5}c$. Notably, SrTiO$_3$ shows a peak of the deceleration factor at nonzero wavevectors, which arises from a slight change of slope in the linear part of the phonon dispersion near the $\Gamma$ point.


\section{Discussion}

The confinement and deceleration factors (which mainly involve the explicit wavevector dependence of the 2D LO phonon/2D phonon polariton) of all three investigated oxide-perovskite compounds are equivalent in magnitude to those of hBN, $C\approx\mathcal{O}(10^{3})$, $D\approx\mathcal{O}(10^{-5})$. In contrast, the propagation quality factors, which are highly dependent on the phonon linewidth through the real and imaginary parts of the optical conductivity, are systematically lower for the oxide perovskites than for hBN. This can straightforwardly be explained by the higher complexity of the 5-atom unit cell of the oxide perovskites compared to the 2-atom unit cell of hBN: the larger number of atoms per unit cell results in a larger number of phonon modes and more possibilities for the high-frequency phonons to relax, which increases the phonon linewidth and accordingly reduces the propagation quality of the 2D phonon polaritons. 



The strong confinement of light in monolayers of oxide perovskites may enable enhanced control over the correlated phases of matter in these materials, for example by interacting with the constrained modes of an optical cavity, which was recently proposed for the paraelectric-to-ferroelectric phase transition bulk and thin-film SrTiO$_3$ and KTaO$_3$ \cite{Ashida2020}. Among the investigated oxide perovskites, LiNbO$_3$ emerges as the most promising candidate for a 2D phonon-polaritonic platform in the terahertz regime, as it combines high confinement and deceleration of light in the high-frequency 2D phonon polaritons with high propagation quality, comparable to that of hBN. The confinement of light in monolayers of LiNbO$_3$ could further enhance optical nonlinearities in this material, which is already being used for optoelectronic applications as thin films and bulk \cite{Wang2018_2,Zhang2019}.


\begin{acknowledgments}
We are grateful to Nicholas Rivera, Yuefeng Nie, Ata\c{c} \.{I}mamo\u{g}lu, and Tobia Nova for useful discussions. This work was supported by the Swiss National Science Foundation (SNSF) under project ID 184259, the DARPA DRINQS Program under award number D18AC00014 (non-equilibrium behavior), and partially by the Photonics at Thermodynamic Limits Energy Frontier Research Center under Grant No. DE-SC0019140 (methods for phonon-polariton calculations). P. N. is a Moore Inventor Fellow through Grant GBMF8048 from the Gordon and Betty Moore Foundation. Calculations were performed at the National Energy Research Scientific Computing Center (NERSC), supported by the Office of Science of the U.S. Department of Energy under Contract No. DE-AC02-05CH11231, and at Harvard's Research Computing Facility.
\end{acknowledgments}


\section*{Appendix A: Computational Details}

We calculate the phonon dispersions, eigenvectors, and Born effective charges from first principles, using the density functional perturbation theory formalism \cite{Gonze1997} as implemented in the \textsc{quantum espresso} code \cite{Giannozzi2009,Giannozzi2017}. We use the \textsc{phono3py} code for the calculations of phonon linewidths \cite{Togo2015}. For the phonon calculations, we converge the Hellmann-Feynman forces to 0.1~$\mu$Ry/bohr and use the 2D implementation of the nonanalytical-term correction for the LO and TO phonons \cite{Sohier2017,Sohier2017_2}. We choose the pseudo-dojo scalar-relativistic optimized norm-conserving Vanderbilt pseudopotentials (ONCVPSP) with valence electron configurations Sr ($4s^2 4p^6 5s^2$), Ti ($3s^2 3p^6 3d^2 4s^2$), K ($3s^2 3p^6 4s^1$), Ta ($5s^2 5p^6 5d^3 6s^2$), Li ($1s^2 2s^1$), Nb ($4s^2 4p^6 4d^4 5s^1$), O ($2s^2 2s^4$), B ($2s^2 2s^1$), and N ($2s^2 2s^3$) \cite{Hamann2013,vansetten2018}. For the exchange-correlation functional, we choose the Perdew-Burke-Ernzerhof revised for solids (PBEsol) form of the generalized gradient approximation (GGA) \cite{csonka:2009}. For the oxide perovskites, we add 14~\ang{} of vacuum along the $c$ direction of each of the 5-atom unit cells, for which we use a plane-wave energy cutoff of 100~Ry, and a 6$\times$6$\times$1 $k$-point mesh to sample the Brillouin zone. For hBN, we use 10~\ang{} of vacuum, 55~Ry energy cutoff, and a 12$\times$12$\times$1 $k$-point mesh, respectively. Crystal structures are visualized using \textsc{vesta} \cite{Momma2011}.


\section*{Appendix B: Lattice optical conductivity}

The frequency-dependent polarization-polarization response of a dielectric material can be written as
\begin{equation}
\mathbf{P}(\omega,\mathbf{q}) = \varepsilon_0 \Pi(\omega,\mathbf{q}) \mathbf{E}(\omega,\mathbf{q})
\end{equation}
where $\mathbf{P}$ is the polarization, $\varepsilon_0$ is the vacuum permittivity, $\Pi$ is the polarization-polarization response tensor, and $\mathbf{E}$ is the electric field. We consider the regime without retardation effects ($q\gg\omega/c)$. $\Pi$ is related to the optical conductivity, $\sigma$, by \cite{Jablan2009,Rivera2019}
\begin{equation}\label{eq:conductivitypolarization}
\sigma(\omega,\mathbf{q})=-i\omega\varepsilon_0\Pi(\omega,\mathbf{q}),
\end{equation}
and the ionic contribution to the polarization-polarization response by infrared-active phonon modes will therefore yield the lattice optical conductivity. In the long-wavelength limit ($q\rightarrow 0$), the ionic contribution to $\Pi$ is given by the well-known expression \cite{Horton1974,Gonze1997}
\begin{equation}
\label{eq:ioniccontribution}
\Pi_{ij}(\omega) = \frac{1}{\Omega} \sum\limits_\alpha \frac{Z_{\alpha,i} Z_{\alpha,j}}{\omega_{\alpha}^2 - \omega^2 - 2i\gamma_{\alpha}\omega},
\end{equation}
where $\Omega$ is the unit-cell volume or area, $\omega_\alpha$ is the eigenfrequency of phonon mode $\alpha$ at the Gamma point, and $\gamma_\alpha$ is the phonon linewidth. Due to the absence of LO-TO splitting in 2D, $\omega_\alpha\equiv\omega_{\alpha,\mathrm{TO}}=\omega_{\alpha,\mathrm{LO}}$. $Z_{\alpha,i}$ is the mode effective charge vector, which is only nonzero for infrared-active phonons and can be written as $\mathbf{Z}_\alpha=\sum_{n} Z^\ast_n \hat{\mathbf{e}}_{\alpha n}/\sqrt{M_n}$, where $Z^\ast_n$ is the Born effective charge tensor of atom $n$, $\hat{\mathbf{e}}_{\alpha n}$ is the phonon eigenvector, and $M_n$ the atomic mass. The numerator of Eq.~(\ref{eq:ioniccontribution}) can be written in vector/tensor form as $Z_{\alpha,i} Z_{\alpha,j}=|\hat{\mathbf{q}}\cdot\mathbf{Z}_\alpha|^2$, and combining Eqs.~(\ref{eq:conductivitypolarization}) and (\ref{eq:ioniccontribution}) yields Eq.~(\ref{eq:opticalconductivity}) from the main text,
\begin{eqnarray}
\sigma(\omega) & = & -i\frac{\omega}{\Omega}\sum\limits_{\alpha} \frac{|\hat{\mathbf{q}}\cdot\mathbf{Z}_{\alpha}|^2}{\omega_\alpha^2 - \omega^2 - 2i\gamma_\alpha\omega} \nonumber\\
& = & -2i\varepsilon_0\omega\sum\limits_{\alpha} \frac{\mathcal{S}_\alpha}{\omega_\alpha^2 - \omega^2 - 2i\gamma_\alpha\omega},
\end{eqnarray}
with $\mathcal{S}_\alpha=|\hat{\mathbf{q}}\cdot{}\mathbf{Z}_\alpha|^2/(2\Omega\epsilon_0)$.



%

\end{document}